# New Confidence Intervals and Bias Comparisons Show that Maximum Likelihood Can Beat Multiple Imputation in Small Samples

Paul T. von Hippel
LBJ School of Public Affairs, University of Texas
2315 Red River, Box Y, Austin, TX 78712
(512) 537-8112
paulvonhippel.utaustin@gmail.com

*Abstract*

When analyzing incomplete data, is it better to use multiple imputation (MI) or full information maximum likelihood (ML)? In large samples ML is clearly better, but in small samples ML's usefulness has been limited because ML commonly uses normal test statistics and confidence intervals that require large samples. We propose small-sample *t*-based ML confidence intervals that have good coverage and are shorter than *t*-based confidence intervals under MI. We also show that ML point estimates are less biased and more efficient than MI point estimates in small samples of bivariate normal data. With our new confidence intervals, ML should be preferred over MI, even in small samples, whenever both options are available.

# 1   INTRODUCTION

Multiple imputation (MI) and full information maximum likelihood (ML) are increasingly popular methods for analyzing data with missing values (Allison, 2002; Little & Rubin, 2002). MI fills in each missing value with a random sample of plausible imputations, while ML integrates the missing values out of the likelihood. The increasing popularity of MI and ML is due in part to their increasing availability in software. For example, the SEM software LISREL, MPlus, AMOS, and SAS PROC CALIS can all use either MI or ML.

When both ML and MI are available, which is better? If the sample is large, the answer is clear: it is better to use ML. One reason for this is that ML is more efficient. In large samples, MI would need an infinite number of imputations to achieve the efficiency of ML (Wang & Robins, 1998). If the number of imputations is small (3 to 10), as it typically is, MI is just slightly less efficient than ML at producing point estimates (Rubin, 1987), but MI is much less efficient at estimating standard errors, confidence intervals, or the fraction of missing information (Graham, Olchowski, & Gilreath, 2007). If the fraction of missing information is large, standard errors estimated by MI can require 200-300 imputed datasets to approach the efficiency of standard errors estimated by ML (Bodner, 2008; Graham et al., 2007). It can be impractical to use so many imputations if the sample is large, or if the analysis runs slowly on each imputed dataset.

With large samples, then, ML is better than MI. But what about small samples? One could get the impression that small samples turn the table and give the advantage to MI. Under

MI it is common to report hypothesis tests and confidence intervals using a *t* statistic whose degrees of freedom decrease with the sample size *n* (Barnard & Rubin, 1999 developed the most popular choice; for alternatives, see Wagstaff & Harel, 2011). By contrast, ML confidence intervals commonly rely on asymptotic normality and so require a large sample. Small-sample *t* intervals have also been proposed for ML, but they are rarely used because they only work for two specific estimands: the mean and the mean difference (Little, 1976, 1988; Morrison, 1973).

In this article, we propose a more general degrees of freedom formula for *t* intervals under ML with missing data. Section 6.2 proposes two very similar formulas; the more general is

$$\hat{v}_{ML} = v_{com}(1 - \hat{\gamma}_{ML})\left(\frac{v_{com} + 1}{v_{com} + 3}\right) \qquad (1).$$

Here $\hat{\gamma}_{ML}$ is an estimate of the fraction of missing information $\gamma$. The estimate $\hat{\gamma}_{ML}$ is calculated directly from the incomplete data using ML, and is more precise than the estimate of $\gamma$ that is commonly used under MI (Savalei & Rhemtulla, 2012). $v_{com}$ is the degrees of freedom that would apply if the data were complete—e.g., $v_{com} = n - 2$ for a simple regression. (Note that $v_{com}$ is the error degrees of freedom which depends on the sample size; it should not be confused with the SEM degrees of freedom which depends on the number of distinct elements in the covariance matrix.)

The formula for $\hat{v}_{ML}$ makes ML a viable approach even in small samples. In simulations using bivariate normal data, we will show that small-sample *t* intervals using $\hat{v}_{ML}$ under ML have good coverage and are shorter and more reliable than the small-sample *t* intervals that are commonly used MI. We will also show that, under both ML and MI, the degrees of freedom estimators perform better if they are constrained to take values no smaller than 3.

In addition to the new confidence intervals, our simulations also evaluate the point estimates produced by ML and MI in small samples of incomplete normal data. We find that MI point estimates are less efficient than ML point estimates in small samples, just as they are in large samples. The reason for MI's lower efficiency is simple: random variation in the imputed values contributes excess variation to the MI estimates.

We also confirm that, while both ML and MI can produce biased point estimates, the biases are larger under MI than they are under ML. Some of these biases have been noticed in previous research, but previous results are scattered and no one has pulled them together to tell a coherent story. We offer an intuitive explanation, as well as a closed-form expression for the bias, which shows when it will be smaller or larger. The pattern of bias is fairly simple. Under MI, the variance of the imputed variable is overestimated (cf. Demirtas, Freels, & Yucel, 2008, Table 2; Yuan, Yang-Wallentin, & Bentler, 2012, Table 2). When the imputed variable is used as a regressor, the estimated regression slope is attenuated (cf. Hoogendoorn, 2009). When the imputed variable is used a regressand, the regression slope is unbiased but the residual variance is overestimated (cf. Chen & Ibrahim, 2013; Kim, 2004). The bias in the residual variance can be eliminated by a change to the Bayesian prior (Kim, 2004; von Hippel, 2013b), but eliminating the other biases is not so easy. Under ML, the biases are similar but considerably smaller.

Overall, with our new confidence intervals, ML estimates can be at least as good as MI in small samples as well as large. As ML is also easier to use than MI, we conclude that ML should be preferred whenever both options are available. However, it is not always the case that both options are available. There are situations when ML is unavailable and MI should be used. We will discuss some of those situations in the Conclusion.

## 2 INCOMPLETE BIVARIATE NORMAL DATA

In this section we introduce the incomplete bivariate normal data that is used in later calculations and simulations. We draw data from an infinite population consisting of standard bivariate normal variables ($X,Y$) with mean and covariance matrix

$$\mu_{XY} = \begin{bmatrix} \mu_X \\ \mu_Y \end{bmatrix} = \begin{bmatrix} 0 \\ 0 \end{bmatrix}$$
$$\Sigma_{XY} = \begin{bmatrix} \sigma_X^2 & \\ \sigma_{XY} & \sigma_Y^2 \end{bmatrix} = \begin{bmatrix} 1 & \\ \rho & 1 \end{bmatrix}$$
(2).

Notice that the regression of $Y$ on $X$ has the same parameter values as the regression of $X$ on $Y$. That is, in the following regression equations,

$$Y = \alpha_{Y.X} + \beta_{Y.X} X + e_{Y.X}, \text{where } e_{Y.X} \sim N(0, \sigma_{Y.X}^2)$$
$$X = \alpha_{X.Y} + \beta_{X.Y} Y + e_{X.Y}, \text{where } e_{X.Y} \sim N(0, \sigma_{X.Y}^2)$$
(3),

the parameter values are $\alpha_{Y.X} = \alpha_{X.Y} = 0$, $\beta_{Y.X} = \beta_{X.Y} = \rho$, and $\sigma_{Y.X}^2 = \sigma_{X.Y}^2 = 1 - \rho^2$.

We define a dummy variable $M$ to indicate whether the value of $Y$ is missing ($M=1$) or observed ($M=0$). Notice that the mean of $M$ is also the proportion $p$ of $Y$ values that are missing. If the variable $M$ is independent of the variables $X$ and $Y$, then values are said to be *missing completely at random* (MCAR). If $M$ depends on $X$ but not on $Y$ (net of $X$), then values are said to be *missing at random* (MAR) (Heitjan & Basu, 1996). There are many ways for values to be MAR. Our simulations focus on the MAR pattern where $Y$ values are *missing if X is negative* (MXN)—i.e., $M = 1$ iff $X < 0$. MXN is a simple but challenging MAR pattern that limits statistical information and yields estimates with large bias in small samples. After we develop an understanding of where the bias comes from, it will be easy to anticipate the bias that would result from other MAR patterns.

It will be helpful to know the conditional moments of $(X,Y)$ given $M$. Let $\mu_{XY.0}$ and $\Sigma_{XY.0}$ be the moments when $M = 0$; likewise let $\mu_{XY.1}$ and $\Sigma_{XY.1}$ be the moments when $M = 1$. If $Y$ values are MCAR, then the conditional moments of $(X,Y)$ given $M$ are the same as the unconditional moments—that is, $\mu_{XY.0} = \mu_{XY.1} = \mu_{XY}$ and $\Sigma_{XY.0} = \Sigma_{XY.1} = \Sigma_{XY}$. But if $Y$ values are MXN then the conditional moments of $(X,Y)$ given $M$ are the moments of a standard bivariate normal distribution where one variable has been truncated at zero. Those moments are (Rose & Smith, 2002, p. 226)

$$\mu_{XY.1} = -\mu_{XY.0} = \sqrt{\frac{2}{\pi}} \begin{bmatrix} 1 \\ \rho \end{bmatrix}$$
$$\Sigma_{XY.1} = \Sigma_{XY.0} = \frac{1}{\pi} \begin{bmatrix} \pi - 2 & \\ (\pi - 2)\rho & \pi - 2\rho^2 \end{bmatrix}$$

(4).

It will also be helpful to know the unconditional moments of $X$, $Y$, $M$, and the interaction $XM$. It is straightforward to calculate those moments using the algebra of expectations:

$$\mu_{XYMXM} = E\begin{bmatrix} X \\ Y \\ M \\ XM \end{bmatrix} = \begin{bmatrix} 0 \\ 0 \\ p \\ p\mu_{X.1} \end{bmatrix} \quad (5).$$

$$\Sigma_{XYMXM} = V\begin{bmatrix} X \\ Y \\ M \\ XM \end{bmatrix} = \begin{bmatrix} 1 & & & \\ \rho & 1 & & \\ p\mu_{X.1} & p\mu_{X.1}\rho & p(1-p) & \\ p\sigma_{X.1}^2 & p\sigma_{X.1}^2\rho & \mu_{X.1}p(1-p) & p(\sigma_{X.1}^2 + \mu_{X.1}^2(1-p)) \end{bmatrix}$$

In interpreting $\Sigma_{XYMXM}$, notice that if $Y$ values are MCAR then $M$ is uncorrelated with $X$ since $\mu_{X.1} = 0$. On the other hand, if $Y$ values are MXN then $M$ is correlated with $X$.

To illustrate the properties of different estimators, we held $p$ and $\rho$ constant at 1/2, we let $Y$ values be MCAR or MXN, and we simulated samples each containing $n$ cases from the population above. Of these $n$ cases, $n_0$ cases were complete, with $M = 0$ and $Y$ observed, and $n_1 = n - n_0$ cases were incomplete, with $M = 1$ and $Y$ missing. For purposes of some formulas below, we suppose that the data are sorted so that the complete cases come first; that is, the observed $Y_i$ values are in cases $i = 1, \ldots, n_0$ and the missing $Y_i$ values are in cases $i = n_0 + 1, \ldots, n$.

It remains only to choose values for the sample size $n$. While there are old rules of thumb suggesting that structural equation models invariably require 100-250 cases, more recent and empirically grounded work shows that the number of cases required depends on the model and data (Hogarty, Hines, Kromrey, Ferron, & Mumford, 2005; MacCallum, Widaman, Zhang, & Hong, 1999; Muthén & Muthén, 2002). In regression analysis, smaller samples can be adequate if the residual variance is small and the collinearity among regressors is low (Kelley & Maxwell, 2003). In factor analysis, smaller samples can be adequate if the model is simple and the communalities are high (Hogarty et al., 2005;

MacCallum et al., 1999). A recent simulation found that, depending on the characteristics of the model and data, the number of cases required to meet certain criteria for power, bias, and solution propriety could be as low as $n=30$ or as high as $n=460$ (Wolf, Harrington, Clark, & Miller, 2013).

To explore the low end of these recommendations, we let $n$ take values of 25 or 100. A sample size of $n=25$ is consistent with past studies on the small-sample properties of missing-data estimators, which have most commonly examined sample sizes down to $n=25$ (Lipsitz, Parzen, & Zhao, 2002; Wagstaff & Harel, 2011), although two studies went as low as $n=10$ (Barnard & Rubin, 1999; von Hippel, 2013b) and two didn't go below $n=40$ (Demirtas et al., 2008; Reiter, 2007). Samples of this size are of practical interest in applied settings where the intended model is simple and it is expensive, impractical, or impossible to gather a large sample. For example, in political science, regressions that compare US states cannot have a sample size greater than n = 50 (Granberg-Rademacker, 2007), and regressions that compare different countries may have n as small as 41 or 46 (Cohen, 2004; Kunovich & Paxton, 2003). The problems of finite populations is an issue in political research, since there are only 50 states in the US and 196 countries in the world. But the use of small samples is not limited to fields with small populations. Pilot clinical trials can use as few as $n = 20$ patients (Barnes, Lindborg, & Seaman, 2006), and the median number of subjects in psychology experiments is just $n = 18$ (Marszalek, Barber, Kohlhart, & Holmes, 2011).

In order to accurately evaluate the properties of the estimators, our research design had many replications for each simulated condition. For conditions with $n=100$, we replicated the simulation 40,000 times; for conditions with *n*=25, we replicated the simulation 160,000 times. The large number of replications ensures that the means and standard deviations of the point estimates, as estimated from the simulation, are usually accurate to two significant digits. To evaluate confidence intervals, we used only 8,000 replications because the calculations were iterative and ran more slowly. With 8,000 replications the mean length of confidence intervals were usually accurate to 1 decimal place, and the estimated coverage of confidence intervals were accurate to within a standard error of 0.25%.

## 3  MAXIMUM LIKELIHOOD (ML)

ML estimation often requires iterative numerical maximization of the likelihood. But in bivariate normal data with values missing at random, ML estimates can be obtained as closed form expressions (Anderson, 1957), which permit us to calculate the biases in closed form as well. Since *X* is complete, ML estimates for the moments of *X* are obtained using the usual complete-data ML formulas:

$$\hat{\mu}_{X,ML} = \frac{1}{n}\sum_{i=1}^{n} X_i$$
$$\hat{\sigma}^2_{X,ML} = \frac{1}{n}\sum_{i=1}^{n}(X_i - \hat{\mu}_{X,ML})^2$$
(6)

And ML estimates for the regression of *Y* on *X* can be obtained simply by ignoring the incomplete cases and applying OLS to the $n_0$ cases with *Y* observed (Anderson, 1957;

Little, 1992; von Hippel, 2007). The ML slope $\hat{\beta}_{Y.X,ML}$ and intercept $\hat{\alpha}_{Y.X,ML}$ are calculated by the usual OLS formulas, and the ML residual variance $\hat{\sigma}^2_{Y.X,ML}$ is calculated by averaging the squared residuals:

$$\hat{\sigma}^2_{Y.X,ML} = \frac{1}{n}\sum_{i=1}^{n}\left(Y_i - \hat{\alpha}_{Y.X,ML} - \hat{\beta}_{Y.X,ML}X_i\right)^2 \qquad (7)$$

However, ML estimates for other parameters cannot be obtained by ignoring the incomplete cases. Instead, because ML estimates are functionally invariant, ML estimates for other parameters can be obtained by transforming the ML estimates that we have already obtained. For example, the following formulas give ML estimates for the mean and variance of *Y* and for the covariance of *Y* with *X*:

$$\begin{aligned}
\hat{\mu}_{Y,ML} &= \hat{\alpha}_{Y.X,ML} + \hat{\beta}_{Y.X,ML}\hat{\mu}_{X,ML} \\
\hat{\sigma}_{XY,ML} &= \hat{\beta}_{Y.X,ML}\hat{\sigma}^2_{X,ML} \\
\hat{\sigma}^2_{Y,ML} &= \hat{\beta}^2_{Y.X,ML}\hat{\sigma}^2_{X,ML} + \hat{\sigma}^2_{Y.X,ML}
\end{aligned} \qquad (8)$$

(Anderson, 1957). Likewise, the following formulas define ML estimates for the regression of *X* on *Y*:

$$\begin{aligned}
\hat{\beta}_{X.Y,ML} &= \hat{\sigma}_{XY,ML}/\hat{\sigma}^2_{Y,ML} \\
\hat{\alpha}_{X.Y,ML} &= \hat{\mu}_{X,ML} - \hat{\beta}_{X.Y,ML}\hat{\mu}_{Y,ML} \\
\hat{\sigma}^2_{X.Y,ML} &= \hat{\sigma}^2_{X,ML} - \hat{\beta}^2_{X.Y,ML}\hat{\sigma}^2_{Y,ML}
\end{aligned} \qquad (9)$$

These estimates follow from the fact that the same formulas define relationships among the parameters.

## 3.1 Bias

The middle rows of Table 1 give the approximate expectations and standard errors of the ML estimates, calculated by taking the mean and standard deviation of the estimates obtained in our simulation. The estimates $\hat{\mu}_{ML}, \hat{\alpha}_{Y.X,ML}, \hat{\beta}_{Y.X,ML}$ are unbiased, while the other estimates have at least slight biases. The biases are worst when the sample is small ($n=25$) and when values are MXN.

The negative biases in the variances $\hat{\sigma}^2_{X,ML}$ and $\hat{\sigma}^2_{Y.X,ML}$ are familiar, and result from the estimates using formulas that divide by $n$ and $n_0$ instead of $n-1$ and $n_0 - 2$. The covariance $\hat{\sigma}^2_{X,ML}$ also has a negative bias which it inherits from $\hat{\sigma}^2_{X,ML}$ through the formula $\hat{\sigma}_{XY,ML} = \hat{\beta}_{Y.X,ML} \hat{\sigma}^2_{X,ML}$.

Less familiar is the positive bias in $\hat{\sigma}^2_{Y,ML}$ that occurs when values are MXN. This has been observed in simulations (Yuan et al., 2012, Table 2), but it has never been explained. To understand the positive bias in $\hat{\sigma}^2_{Y,ML}$, consider the definition $\hat{\sigma}^2_{Y,ML} = \hat{\beta}^2_{Y.X,ML} \hat{\sigma}^2_{X,ML} + \hat{\sigma}^2_{Y.X,ML}$ and notice that the square $\hat{\beta}^2_{Y.X,ML}$ is positively biased, even though $\hat{\beta}_{Y.X,ML}$ itself is unbiased. A simple expression for the bias in $\hat{\beta}^2_{Y.X,ML}$ is

$$Bias(\hat{\beta}^2_{Y.X,ML}) = E(\hat{\beta}^2_{Y.X,ML}) - \beta^2_{Y.X} = V(\hat{\beta}_{Y.X,ML}) \tag{10},$$

which follows immediately from the identity $V(\hat{\beta}_{Y.X,ML}) = E(\hat{\beta}^2_{Y.X,ML}) - [E(\hat{\beta}_{Y.X,ML})]^2$. When values are MXN, the positive bias in $\hat{\beta}^2_{Y.X,ML}$ outweighs the negative biases in $\hat{\sigma}^2_{X,ML}$ and $\hat{\sigma}^2_{Y.X,ML}$, so that the net bias in $\hat{\sigma}^2_{Y,ML}$ is positive, and large when the sample is small.

Why is the bias in $\hat{\beta}^2_{Y.X,ML}$ larger with a small sample size or an MXN pattern? It is because the bias is equal to $V(\hat{\beta}_{Y.X,ML})$, which is larger when we reduce the sample size or restrict the range of observed $X$ values that are used in the regression of $Y$ on $X$. The MXN pattern is one way to restrict the range of $X$.

More explicitly, the asymptotic covariance matrix of the regression estimates is

$$\Sigma_{\hat{\alpha}\hat{\beta}_{Y.X,ML}} = V\begin{pmatrix}\hat{\alpha}_{Y.X,ML} \\ \hat{\beta}_{Y.X,ML}\end{pmatrix} = \frac{\sigma^2_{Y.X}}{n_0}\begin{bmatrix} 1 + \mu^2_{X.0}/\sigma^2_{X.0} & \\ -\mu_{X.0}/\sigma^2_{X.0} & 1/\sigma^2_{X.0}\end{bmatrix} \tag{11},$$

(Little & Rubin, 2002), which is larger when values are MXN since the MXN condition implies that $\sigma^2_{X.0}$ is small and $\mu^2_{X.0}$ is large. More generally, $Bias(\hat{\beta}^2_{Y.X,ML}) = V(\hat{\beta}_{Y.X,ML})$ will be larger under any MAR pattern that yields a small $\sigma^2_{X.0}$, while the MCAR pattern causes relatively little bias since $\sigma^2_{X.0} = \sigma^2_X$.

We can now understand the potential for bias in the regression of $X$ on $Y$. When $n=25$ and values are MXN, the slope $\hat{\beta}_{X.Y,ML} = \hat{\sigma}_{XY,ML}/\hat{\sigma}^2_{Y,ML}$ has a negative bias since its numerator has a negative bias and its denominator has a positive bias. The bias of the intercept is opposite to that of the slope; this follows from the definition of the intercept as $\hat{\alpha}_{X.Y,ML} = \hat{\mu}_{Y,ML} - \hat{\beta}_{X.Y,ML}\hat{\mu}_{X,ML}$. The residual variance $\hat{\sigma}^2_{X.Y,ML} = \hat{\sigma}^2_{X,ML} - \hat{\beta}^2_{X.Y,ML}\hat{\sigma}^2_{Y,ML}$ has a negative bias that comes partly from negative bias in $\hat{\sigma}^2_{X,ML}$, but mostly from positive bias in $\hat{\beta}^2_{X.Y,ML}$. An argument like that in equation (10) shows that $Bias(\hat{\beta}^2_{X.Y,ML}) = V(\hat{\beta}_{X.Y,ML})$.

## 4   MAXIMUM LIKELIHOOD MULTIPLE IMPUTATION (MLMI)

Values can be imputed conditionally on the ML estimates (Wang & Robins, 1998), a practice that we call *maximum likelihood multiple imputation* (MLMI). MLMI is not the most common type of MI, but its simplicity will ease our calculations and serve as a bridge to the more common type of MI.

It is helpful to start with a simplified situation where values are imputed just once—a process that we call maximum likelihood *single* imputation (MLSI). In MLSI, we fill in missing $Y$ values by regression on $X$, where the coefficients of the regression model are ML estimates:

$$Y_{MLSI,i} = \hat{\alpha}_{Y.X,ML} + \hat{\beta}_{Y.X,ML} X_i + e_{MLSI,i}, i = n_1 + 1, \dots, n \qquad (12)$$
$$\text{where } e_{MLSI,i} \sim N(0, \hat{\sigma}^2_{Y.X,ML})$$

The result is a partly observed, partly imputed variable $Y_{obsMLSI}$ consisting of $n_0$ observed values $Y_{obs}$ drawn from equation (3) and $n_1$ imputed values $Y_{MLSI}$ drawn from equation (12). MLSI estimates of means, variances, covariances, and regressions are obtained by applying the usual complete-data formulas to the partly imputed data.[1]

In MLMI, the process of imputation and estimation is repeated $D$ times with each repetition yielding a new MLSI estimate—e.g., $\hat{\mu}_{MLSI,d}$, $d = 1,\dots D$. The MLSI estimates are averaged to yield MLMI estimates—e.g., $\hat{\mu}_{MLMI} = D^{-1} \sum_{d=1}^{D} \hat{\mu}_{MLSI,d}$

---

[1] In our simulation we calculated variances and covariances from imputed data by dividing each sum of squares by its degrees of freedom. Other divisors are also possible.

Since averaging does not affect the expectation of an estimator, the bias of the MLMI estimators is the same as the bias of the MLSI estimators. When discussing bias, we will focus on the MLSI estimates, which makes the mathematics a little simpler.

### 4.1 Bias

The middle rows of Table 1 give the expectations and standard errors of the MLMI estimates, as calculated from our simulation with $D = 5$ imputations. The standard errors of the MLMI estimates are slightly larger than those of the ML estimates—a result that accords with large-sample theory (Wang & Robins, 1998).

The biases of the MLMI estimates are also similar to those of the ML estimates, through slightly larger. In particular, when values are MXN, $\hat{\sigma}^2_{Y,MLSI}$ has a positive bias which is large when $n = 25$. The positive bias in $\hat{\sigma}^2_{Y,MLSI}$ is responsible for the negative bias in $\hat{\beta}_{X.Y,MLSI} = \hat{\sigma}_{XY,MLSI}/\hat{\sigma}^2_{X,MLSI}$, which in turn is responsible for the positive bias in $\hat{\alpha}_{X.Y,MLSI} = \hat{\mu}_{Y,MLSI} - \hat{\beta}_{Y.X,MLSI}\hat{\mu}_{X,MLSI}$. When values are MCAR, on the other hand, the bias of $\hat{\sigma}^2_{Y,MLSI}$ turns slightly negative.

To understand the biases of the MLSI estimates, it is helpful to summarize the distribution of $Y_{obsMLSI}$ in a single equation. Using the following symbols for the errors of the ML estimates,

$$\begin{aligned} \Delta\hat{\alpha}_{Y.X,ML} &= \hat{\alpha}_{Y.X,ML} - \alpha_{Y.X} \\ \Delta\hat{\beta}_{Y.X,ML} &= \hat{\beta}_{Y.X,ML} - \beta_{Y.X} \\ \Delta\hat{\sigma}^2_{Y.X,ML} &= \hat{\sigma}^2_{Y.X,ML} - \sigma^2_{Y.X} \end{aligned} \quad (13),$$

we can summarize the distribution of $Y_{obsMLSI}$ using a single regression equation:

$$Y_{obs\ MLSI} = \alpha + M\Delta\hat{\alpha}_{Y.X,ML} + \beta X + XM\Delta\hat{\beta}_{Y.X,ML} + e_{obsMLSI},$$
$$\text{where } e_{obsMLSI} \sim N(0, \sigma^2_{Y.X} + M\Delta\hat{\sigma}^2_{Y.X,ML}) \tag{14}$$

In the fully observed cases, this reduces to regression (3). In the imputed cases, it reduces to regression (12).

In equation (14), the regressors *M* and *XM* contribute extra variance to *Y*, and that extra variance increases with the coefficients $\Delta\hat{\alpha}_{Y.X,ML}$ and $\Delta\hat{\beta}_{Y.X,ML}$. So when $\Delta\hat{\alpha}_{Y.X,ML}$ and $\Delta\hat{\beta}_{Y.X,ML}$ are large—that is, when $\hat{\alpha}_{Y.X,ML}$ and $\hat{\beta}_{Y.X,ML}$ are variable—the variance of $Y_{obsMLSI}$ has positive bias even though the bias of $\hat{\sigma}^2_{Y.X,ML}$ is negative.

More explicitly, a calculation in the Appendix shows that the bias of $\hat{\sigma}^2_{Y,MLSI}$ is approximately

$$Bias(\hat{\sigma}^2_{Y,MLSI}) = tr\left(\Sigma_{MXM}\Sigma_{\hat{\alpha}\hat{\beta}_{Y.X,ML}}\right) + Bias(\hat{\sigma}^2_{Y.X,ML})\,p \tag{15},$$

where $\Sigma_{MXM}$ is the covariance matrix of (*M*,*XM*)—that is, the lower right corner of the big matrix in (5)—and $\Sigma_{\hat{\alpha}\hat{\beta}_{Y.X,ML}}$ is the covariance matrix of the ML regression estimates, from (11).

The first component in the bias is positive, but the second component is negative because $\hat{\sigma}^2_{Y.X,ML}$ is a negatively biased estimator of $\sigma^2_{Y.X}$. When values are MXN, the first bias component is relatively large so the net bias is positive. When values are MCAR, the first bias component is relatively small so the net bias is negative. These patterns are plainly visible in Table 1.

## 5   POSTERIOR DRAW MULTIPLE IMPUTATION (PDMI)

MLMI is not the most popular form of imputation. Instead, the most popular form of imputation fills in missing values with draws from the posterior predictive distribution. In most accounts, this approach is known simply as multiple imputation (MI) (Rubin, 1987), but here we call it *posterior draw multiple imputation* (PDMI) in order to distinguish it from MLMI.

Again it is helpful to start with a simplified situation where values are imputed just once—i.e., posterior draw *single* imputation (PDSI). Under PDSI, we fill in missing *Y* values by regression on *X*, much as we do under MLSI:

$$Y_{PDSI,i} = \hat{\alpha}_{Y.X,PD} + \hat{\beta}_{Y.X,PD} X_i + e_{PDSI,i}, i = n_1 + 1 \ldots n \quad (16)$$
$$\text{where } e_{PDSI,i} \sim N(0, \hat{\sigma}^2_{Y.X,PD})$$

Under PDMI, however, the estimates $\hat{\alpha}_{Y.X,PD}, \hat{\beta}_{Y.X,PD}, \hat{\sigma}^2_{Y.X,PD}$ used in the imputation model are not ML estimates but posterior draw (PD) estimates drawn at random from the posterior distribution of the parameters given the observed data.

The PD estimates are obtained as follows (Little & Rubin, 1989). First, $\hat{\sigma}^2_{Y.X,PD}$ is drawn from an inverse-chi-square distribution that is scaled by $\hat{\sigma}^2_{Y.X,ML}$,

$$\hat{\sigma}^2_{Y.X,PD} = \hat{\sigma}^2_{Y.X,ML} \frac{n_0}{U}, \text{where } U \sim \chi^2_{n_0-2+\nu_{prior}} \quad (17),$$

Here $\nu_{prior}$ is the prior degrees of freedom (Kim, 2004), which is set in the imputation software. The most widely available choice in software is $\nu_{prior} = 0$, which corresponds to the Jeffreys prior.

Next, $\hat{\alpha}_{Y.X,PD}, \hat{\beta}_{Y.X,PD}$ are drawn from a bivariate normal distribution whose covariance matrix is just the ML covariance matrix rescaled using $\hat{\sigma}^2_{Y.X,PD}$:

$$\begin{bmatrix} \hat{\alpha}_{Y.X,PD} \\ \hat{\beta}_{Y.X,PD} \end{bmatrix} \sim N\left( \begin{bmatrix} \hat{\alpha}_{Y.X,ML} \\ \hat{\beta}_{Y.X,ML} \end{bmatrix}, \Sigma_{\hat{\alpha}\hat{\beta}_{Y.X,ML}} \frac{\hat{\sigma}^2_{Y.X,PD}}{\hat{\sigma}^2_{Y.X,ML}} \right) \qquad (18)$$

Notice that the PD estimates $\hat{\alpha}_{Y.X,PD}, \hat{\beta}_{Y.X,PD}$, though unbiased, are about twice as variable as the ML estimates $\hat{\alpha}_{Y.X,ML}, \hat{\beta}_{Y.X,ML}$. The PD estimates have all the variability of the ML estimates, plus a similar amount of variability that is added by the posterior draws:

$$\Sigma_{\hat{\alpha}\hat{\beta}_{Y.X,PD}} = V \begin{bmatrix} \hat{\alpha}_{Y.X,PD} \\ \hat{\beta}_{Y.X,PD} \end{bmatrix} = \Sigma_{\hat{\alpha}\hat{\beta}_{Y.X,ML}} + \Sigma_{\hat{\alpha}\hat{\beta}_{Y.X,ML}} \frac{\hat{\sigma}^2_{Y.X,PD}}{\hat{\sigma}^2_{Y.X,ML}} \qquad (19)$$
$$\approx 2\Sigma_{\hat{\alpha}\hat{\beta}_{Y.X,ML}}$$

As under MLSI, under PDSI estimates are obtained by applying the usual complete-data formulas to the partly imputed data. As under MLMI, under PDMI estimates are obtained by averaging PDSI estimates across *D* imputations. More explicitly, in each iteration $d = 1,\ldots D$, we draw new parameter estimates $\hat{\alpha}_{Y.X,PD,d}, \hat{\beta}_{Y.X,PD,d}, \hat{\sigma}^2_{Y.X,PD,d}$ from the posterior, we re-impute the missing values, and we calculate new PDSI estimates—e.g., $\hat{\mu}_{PDSI,d}$—from the imputed data. We then average the PDSI estimates to obtain PDMI estimates—e.g., $\hat{\mu}_{PDMI} = D^{-1} \sum_{d=1}^{D} \hat{\mu}_{PDSI,d}$.

## 5.1 Bias

Table 1 gives the expectations and standard errors of the PDMI estimates, as calculated from the simulation with $D = 5$ imputations. The standard errors of the PDMI estimates are larger than those of the MLMI estimates—a result which accords with large-sample theory (Wang & Robins, 1998). The biases of PDMI affect the same parameters as the biases of MLMI, but the biases of PDMI are considerably larger. In discussing the biases, we will focus on the PDSI estimates, which have the same biases as the PDMI estimates but are mathematically simpler.

A striking difference between the PDSI and MLSI biases is that under PDSI the bias of $\hat{\sigma}^2_{Y,PDSI}$ is positive not just when values are MXN but when values are MCAR as well. The positive bias in $\hat{\sigma}^2_{Y,PDSI}$ is responsible for the negative bias in $\hat{\beta}_{X.Y,PDSI} = \hat{\sigma}_{XY,PDSI}/\hat{\sigma}^2_{X,PDSI}$, which in turn is responsible for the positive bias in $\hat{\alpha}_{X.Y,PDSI} = \hat{\mu}_{Y,PDSI} - \hat{\beta}_{Y.X,PDSI}\hat{\mu}_{X,PDSI}$.

Using the same approximation that we used for MLSI (see Appendix), we get the following expression for the asymptotic bias of $\hat{\sigma}^2_{Y,PDSI}$:

$$Bias(\hat{\sigma}^2_{Y,PDSI}) = tr\left(\Sigma_{MXM}\Sigma_{\hat{\alpha}\hat{\beta}_{Y.X,PD}}\right) + Bias(\hat{\sigma}^2_{Y.X,PD})\,p \qquad (20)$$

The first component in the bias is positive, and the second component is also positive because the estimate $\hat{\sigma}^2_{Y.X,PD}$ has a bias of

$$\begin{aligned} Bias(\hat{\sigma}^2_{Y.X,PD}) &= E(\hat{\sigma}^2_{Y.X,PD}) - \sigma^2_{Y.X} \\ &= \frac{2 - \nu_{prior}}{(n_0 - 2) - (2 - \nu_{prior})}\sigma^2_{Y.X} \end{aligned} \qquad (21)$$

which is positive under the usual choice $v_{prior} = 0$ (Kim, 2004; von Hippel, 2013b).

Because both components of $Bias(\hat{\sigma}^2_{Y,PDSI})$ are positive, the net bias in $\hat{\sigma}^2_{Y,PDI}$ is positive even when values are MCAR. In addition, the first bias component is about twice as large under PDSI as it is under MLSI, because $\Sigma_{\hat{\alpha}\hat{\beta}_{Y.X,PD}} \approx 2\Sigma_{\hat{\alpha}\hat{\beta}_{Y.X,ML}}$. That is why the biases of PDSI are worse than they are under MLSI: the variation in the posterior draws inadvertently adds to the variation in $Y$.

## 5.2 Root mean squared error

Table 3 evaluates the bias-variance tradeoff by comparing the estimators with respect to root mean squared errors (RMSE). For each sample size and pattern, gives the RMSE of each estimator and ranks the estimators from 1 (smallest RMSE) to 6 (largest RMSE).

Across all the estimands, the smallest RMSEs are typically achieved by the ML estimates, followed by the MLMI estimates and then the PDMI estimates with $v_{prior} = 7$. Exceptions to this pattern can occur when an estimate is biased toward zero. Biasing an estimate toward zero can sometimes produce a favorable bias-variance tradeoff, although such a tradeoff is by no means guaranteed (Draper & Nostrand, 1979; Tibshirani, 1996).

For example, in estimating $\sigma^2_{Y.X}$, the PDMI estimator with $v_{prior} = 7$ achieves the best RMSE with an estimate that is biased toward zero. Similarly, in estimating $\beta_{X.Y}$, the PDMI estimator with $v = -2$ is achieves the best RMSE with an estimate that is biased toward zero. Overall, though, it is impossible to recommend the PDMI estimator with $v_{prior} = -2$ because it has the worst or second-worst RMSE in estimating 7 of the 9 tabled parameters.

Indeed, for some parameters, its RMSE is not just large but undefined. These undefined RMSEs show up as very large entries in the table. These are not typos.

ML turns in the best RMSEs overall. At *n*=25, ML has the best RMSE for 6 of the 9 estimands. And for some estimands where ML doesn't have the best RMSE, we can see its ranking improve as the sample size increases from *n*=25 to *n*=100. This is not surprising. As the sample size increases, asymptotic theory tells us that all the biases vanish and the smallest RMSE will be achieved by the ML estimates, followed by the MLMI estimates, followed by the PDMI estimates (Wang & Robins, 1998). In addition, the differences among the PDMI estimators vanish in large samples as the choice of prior become immaterial.

### 5.3 Bias reduction

A simple way to reduce the bias of PDMI estimates is to increase $v_{prior}$. Table 2 illustrates the effect of $v_{prior}$ on simulated PDMI results with $D = 5$ imputations and $v_{prior} =$ –2, 0, 2, or 7. The choice $v_{prior} = 2$ eliminates the bias in $\hat{\sigma}^2_{Y.X,PDMI}$ (Kim 2004; see also our equation (21)) and reduces bias in other estimates by reducing $\Sigma_{\hat{\alpha}\hat{\beta}_{Y.X,PD}}$. The choice $v_{prior} = 7$ further reduces bias in most estimates by reducing $\Sigma_{\hat{\alpha}\hat{\beta}_{Y.X,PD}}$. Although $v_{prior} = 7$ yields a negatively biased estimate $\hat{\sigma}^2_{Y.X,PDMI}$ the bias comes with a greater reduction in variability so that the estimate $\hat{\sigma}^2_{Y.X,PDMI}$ has the smallest mean squared error that can be achieved by any choice of $v_{prior}$ (von Hippel, 2013b).

Unfortunately, the choices $v_{prior} = 2$ and 7 are not available in popular commercial software. Instead, Stata's *mi impute* command defaults to $v_{prior} = 0$ (the Jeffreys prior) which yields bias, and offers as alternative $v_{prior} = -2$ (the uniform prior) which makes the bias worse. $v_{prior} = -2$ worsens the bias in two way—first by increasing the $Bias(\hat{\sigma}^2_{Y.X,PD})$ according to equation (21), and second by increasing $\Sigma_{\hat{\alpha}\hat{\beta}_{Y.X,PD}}$. In fact, with $v_{prior} = -2$ the bias and standard error of some estimates can be not just large but *undefined*. More specifically, with $v_{prior} = -2$, $\hat{\sigma}^2_{Y.X,PDMI}$ and $\hat{\sigma}^2_{Y,PDMI}$ have undefined standard errors occur when there are $n_0 \leq 7$ observed values (von Hippel, 2013b), which is a condition that occasionally occurred in the simulation. In the simulation, undefined expectations and standard errors show up as very large values in Table 2. These values are so large that they look like typos—but they are not.

## 6 CONFIDENCE INTERVALS

In analyzing incomplete data, we typically want not just a point estimate but a confidence interval or equivalent hypothesis test. Given the smaller bias and greater precision of ML estimates, there is clear potential for ML confidence intervals to be shorter than PDMI confidence intervals with similar coverage. In small samples, however, confidence intervals have been more fully developed for PDMI than for ML. In this section, we review the commonly used PDMI confidence intervals and then propose new ML confidence intervals to compete with them. (We are pursuing the estimation of MLMI confidence intervals separately and will not discuss them here.)

## 6.1 PDMI confidence intervals

PDMI confidence intervals are typically calculated as

$$\hat{\theta}_{PDMI} \pm t\hat{V}_{PDMI}^{1/2} \qquad (22).$$

where $\hat{\theta}_{PDMI}$ is a parameter estimate, $\hat{V}_{PDMI}$ is the estimated variance of the parameter estimate, and $t$ is a quantile from a $t$ distribution[2] with $\hat{v}_{PDMI}$ degrees of freedom. The degrees of freedom are estimated as

$$\hat{v}_{PDMI} = \left(\frac{1}{\hat{v}_D} + \frac{1}{\hat{v}_{obs}}\right)^{-1} \qquad (23)$$

(Barnard & Rubin, 1999), where

$$\begin{aligned}\hat{v}_D &= (D-1)\hat{\gamma}_{PDMI}^{-2} \\ \hat{v}_{obs} &= v_{com}(1 - \hat{\gamma}_{PDMI})\left(\frac{v_{com}+1}{v_{com}+3}\right)\end{aligned} \qquad (24)$$

are the respective degrees of freedom in the imputations and in the observed data. Here $\hat{\gamma}_{PDMI}$ is an estimate of the fraction of missing information, and $v_{com}$ is the degrees of freedom that would be used if the data were complete—e.g., $v_{com} = n - 2$ for a simple linear regression. Finally, $\hat{\gamma}_{PDMI}$ and $\hat{V}_{PDMI}$ are estimated as follows:

$$\begin{aligned}\hat{V}_{PDMI} &= \widehat{W}_{PDMI} + \left(1 + \frac{1}{D}\right)\hat{B}_{PDMI} \\ \hat{\gamma}_{PDMI} &= \left(1 + \frac{1}{D}\right)\frac{\widehat{W}_{PDMI}}{\hat{V}_{PDMI}}\end{aligned} \qquad (25)$$

(Rubin, 1987), where

---

[2] The $t$ distribution is just an approximation here. The true small-sample distribution of PDMI estimates is not exactly $t$ (von Hippel, 2013b; Wagstaff & Harel, 2011).

$$\widehat{W}_{PDMI} = \frac{1}{D} \sum_{d=1}^{D} \widehat{W}_{PDSI,d}$$
$$\widehat{B}_{PDMI} = \frac{1}{D-1} \sum_{d=1}^{D} (\hat{\theta}_{PDSI} - \hat{\theta}_{PDMI})^2 \tag{26}$$

are the variances within and between the $D$ imputed datasets. More specifically, $\widehat{B}_{PDMI}$ is the variance of $\hat{\theta}_{PDSI,d}$ from one imputed dataset to another, and $\widehat{W}_{PDMI}$ is the average value of the statistic $\widehat{W}_{PDSI,d}$ that would estimate the variance of $\hat{\theta}_{PDSI,d}$ if the SI dataset were complete rather than partly imputed (e.g., $\widehat{W}_{PDSI,d} = \hat{\sigma}^2_{Y,PDMI}/n$ if $\hat{\theta}_{PDSI,d} = \hat{\mu}_{Y,DMI,d}$).

In initial simulations, we found that estimation error in $\hat{v}_{PDMI}$ could occasionally result in confidence intervals that were orders of magnitude longer than they needed to be to achieve nominal coverage. Because of estimation error, it is quite possible for the estimated degrees of freedom $\hat{v}_{PDMI}$ to be close to 0, although it is hard to imagine a realistic situation where the true degrees of freedom would be less than 2 or 3. Small values of $\hat{v}_{PDMI}$ are a serious problem because as $\hat{v}_{PDMI}$ approaches zero, the $t$ quantile used in the confidence interval becomes arbitrarily large. For example, if $\hat{v}_{PDMI} = 1$, a 95% confidence interval has $t \approx 12.7$; if $\hat{v}_{PDMI} = 1/2$, a 95% confidence interval has $t \approx 165$, and if $\hat{v}_{PDMI} = 1/4$, a 95% confidence interval has $t \approx 43{,}640$. To put this another way, when $\hat{v}_{PDMI} \leq 2$ the $t$ distribution has such heavy tails that its variance is undefined, and when $\hat{v}_{PDMI} \leq 1$ the mean of the $t$ distribution is undefined as well. Small $\hat{v}_{PDMI}$ values occurred with some frequency in our simulation; for example, with $n=25$ and values

MXN, about 24% of $\hat{v}_{PDMI}$ values were less than 3, 11% were less than 2, and 2% were less than 1.

To ensure against too-small values of $\hat{v}_{PDMI}$, we replaced $\hat{v}_{PDMI}$ with the bounded estimator

$$\tilde{v}_{PDMI} = \max(3, \hat{v}_{PDMI}) \tag{27}$$

which, since it never goes under 3, can never produce a *t* statistic with an undefined mean, an undefined variance, or very extreme quantiles. This change will have little impact on coverage if in fact the true degrees of freedom are at least 3—which is almost always a safe bet. If for some reason one believes that the true degrees of freedom are less than 3, one can use a lower bound of 2 or 2.5 which in our simulation produced similar results.

## 6.2 ML confidence intervals

Under ML, the standard error of a scalar parameter estimate $\hat{\theta}_{ML}$ can be estimated as $\hat{I}_{obs,i}^{-1/2}$, which is the square root of a diagonal element of the inverse of the observed information matrix $\hat{I}_{obs}$. Then in large samples, a normal confidence interval can be calculated as

$$\hat{\theta}_{ML} \pm z\hat{I}_{obs,i}^{-1/2} \tag{28}$$

where $z$ is a quantile from the standard normal distribution.

In small samples, normal confidence intervals are too short to achieve full coverage, and better results can be achieved with an approximate[3] *t* interval like

$$\hat{\theta}_{ML} \pm t \hat{I}_{obs,i}^{-1/2} \tag{29}$$

The only trick is estimating the degrees of freedom for the *t* quantile. Some degrees of freedom estimators have been proposed, but they are limited to specific estimates such as $\hat{\mu}_{Y,ML}$ or $\hat{\mu}_{Y,ML} - \hat{\mu}_{X,ML}$ (Little, 1976, 1988; Morrison, 1973).

We now propose a more general estimator for degrees of freedom under ML. There are two ways to reach similar results. For the first approach, consider estimating a regression. In complete data, confidence intervals for the regression slopes and intercept would be calculated using a *t* distribution with degrees of freedom

$$\nu_{com} = n - k \tag{30}$$

where *n* is the sample size *k* is the number of regression parameters (slopes and intercept).

With incomplete data, we can use the same formula but replace *n* with the "effective sample size," by which we mean the sample size that, in complete data, would provide the same amount of information about the parameter as we get from the incomplete data. The effective sample size in incomplete data is estimated by $n(1 - \hat{\gamma}_{ML})$ which is the sample size reduced by an estimate $\hat{\gamma}_{ML}$ of the fraction of missing information. So the degrees of freedom for a regression parameter is

---

[3] Again, the *t* distribution is just an approximation here. The true small-sample distribution of ML estimates with incomplete data is not exactly *t* (Little & Rubin, 2002; von Hippel, 2013b)

$$\hat{v}_{ML}^* = n(1 - \hat{\gamma}_{ML}) - k \tag{31}$$

This formula can be used whenever we estimate a linear regression. Outside of the regression context, though, it is not always the case that the complete data degrees of freedom $v_{comp}$ is estimated simply by subtracting the number of parameters from the sample size. We therefore need a more general formula that will apply whenever where $v_{comp}$ is known. To motivate this more general formula, remember that, in large samples, the ML estimator is equivalent to the PDMI estimator with infinite imputations (Wang & Robins, 1998). This is a large-sample relationship, but it remains useful, if approximate, in small samples as well. Exploiting the relationship we take the limit of $\hat{v}_{PDMI}$ as the number of imputations $D$ goes to infinity, and arrive at the following estimate for the degrees of freedom under ML:

$$\hat{v}_{ML} = v_{com}(1 - \hat{\gamma}_{ML})\left(\frac{v_{com} + 1}{v_{com} + 3}\right) \tag{32}$$

$\hat{v}_{ML}$ and $\hat{v}_{ML}^*$ are not equal but they are very similar. For example, in a simple linear regression with $n=25$, $k=2$, and $\hat{\gamma}_{ML} = .5$, the two formulas give $\hat{v}_{ML} = 10.6$ and $\hat{v}_{ML}^* = 10.5$.

To calculate $\hat{v}_{ML}$ or $\hat{v}_{ML}^*$, we need $\hat{\gamma}_{ML}$ which is an ML estimate of the fraction of missing information. Following Savalei and Rhemtulla (2012) we calculate $\hat{\gamma}_{ML}$ in two steps. First, we enter the incomplete data into ML software to obtain ML point estimates $\hat{\mu}_{XY,ML}, \hat{\Sigma}_{XY,ML}$ and an estimate $\hat{I}_{obs}$ of the observed information. Then we supply the same software with the point estimates $\hat{\mu}_{XY,ML}, \hat{\Sigma}_{XY,ML}$ and a sample size $n$ which implies that the data are complete. This tricks the software into calculating $\hat{I}_{com}$, which estimates of the information that would be available if the data were complete.

For each estimand, the fraction of information that is observed is $\hat{I}_{obs,i}^{-1}\hat{I}_{com,i}$, so the fraction of missing information is

$$\hat{\gamma}_{ML} = 1 - \hat{I}_{obs,i}^{-1}\hat{I}_{com,i} \tag{33}$$

Because of estimation error in $\hat{\gamma}_{ML}$, it is possible though very rare for $\hat{v}_{ML}$ or $\hat{v}_{ML}^*$ to take values less than 3, which as discussed earlier will leave the *t* distribution with an undefined variance and confidence intervals with unnecessary length. To avoid this, we recommend setting a lower bound of 3 on the degrees of freedom—i.e., using

$$\begin{aligned}\tilde{v}_{ML} &= \max(3, \hat{v}_{ML})\\ \text{or } \tilde{v}_{ML}^* &= \max(3, \hat{v}_{ML}^*)\end{aligned} \tag{34}$$

which mirrors the formula that we offered for $\tilde{v}_{PDMI}$. N

In an online Appendix, we illustrate how to calculate $\tilde{v}_{ML}$ and $\hat{\gamma}_{ML}$ using the CALIS procedure in SAS software. Savalei and Rhemtulla (2012) show how to calculate $\hat{\gamma}_{ML}$ in MPlus, R, and EQS. It would be easy for vendors to modify any of these software packages to provide $\hat{\gamma}_{ML}$ and $\tilde{v}_{ML}$ or $\tilde{v}_{ML}^*$ automatically.

### 6.3 Results

In our simulation, we estimated the mean length and coverage of nominal 95% confidence intervals. For most of the parameters ($\alpha_{Y.X}, \beta_{Y.X}, \mu_Y, \alpha_{X.Y}, \beta_{X.Y}, \sigma_{XY}$) we applied the confidence interval formulas directly to the parameter estimates. For the variance parameters ($\sigma^2_{X.Y}$, $\sigma^2_Y$, $\sigma^2_{Y.X}$) we used a cube-root transformation to bring the estimates closer to a normal distribution (Hawkins & Wixley, 1986).[4]

Table 4 gives the mean length and coverage of confidence intervals using the usual formulas; that is, under ML we used the normal confidence intervals in (28) and under PDMI we used *t*-based confidence intervals with the unbounded degrees of freedom estimator in (23). The results highlight the limitations of the usual formulas. Under ML the confidence intervals are too short and cover the estimands less than 95% of the time. With n=100 the coverage is just 0-3% under the nominal level, but with n=25 is 5-6% under the nominal level for the regression intercept and slope, and 8-14% under the nominal level for the residual variance. Under PDMI, the confidence intervals have closer to 95% coverage but are longer, on average, than they need to be. In fact, with *n*=25 some of the mean lengths estimated from the PDMI simulation are so large—e.g., 2.6E+31—that the true mean length must be not just large but undefined. An undefined mean length can occur, as noted earlier, if $\hat{v}_{PDMI} \leq 1$.

Table 5 displays results obtained from our improved confidence intervals. Under PDMI we used *t*-based confidence intervals with $\tilde{v}_{PDMI}$ degrees of freedom, and under ML we used *t*-based confidence intervals with $\tilde{v}^*_{ML}$ degrees of freedom. (Very similar results, not shown, were obtained with $\tilde{v}_{ML}$.) Remember that $\tilde{v}_{PDMI}$, $\tilde{v}_{ML}$, and $\tilde{v}_{ML}^*$ are all constrained to take values no smaller than 3.

---

[4] More specifically, for the untransformed parameters we obtained point estimates and standard errors using the usual formulas. We then cube-root transformed the parameter estimates, and used the delta rule to transform the standard error. We calculated confidence intervals using the transformed parameter estimates and standard errors, then applied the inverse transformation (the cube) to the endpoints of the confidence intervals. Note that Little and Rubin (2002) suggest using a log transformation for a similar purpose. However, a log transformation does a relatively poor job of normalizing variance estimates (Hawkins & Wixley, 1986), and produced poor (unnecessarily long) confidence intervals in our initial simulations.

The results highlight our improvements. The PDMI confidence intervals still have close to 95% coverage but they are shorter than they were before, and they always have defined mean lengths. The fact that the PDMI confidence intervals have close to 95% coverage is a little surprising in view of the bias in the point estimates. Evidently PDMI confidence intervals are long enough to cover not just the variability of the point estimate, but to cover some bias as well.

The ML confidence intervals are also improved. They are longer than they were before and the extra length brings them within 2% of nominal coverage for most parameters. At n=25, coverage is a bit lower for the parameters whose point estimates are biased. Evidently the ML confidence intervals are only long enough to cover variability in the point estimate. Unlike the PDMI confidence intervals, the ML confidence intervals are not long enough to cover bias as well.

On the whole, the ML confidence intervals are quite competitive with the PDMI confidence intervals. The ML confidence intervals have good coverage and are shorter than the PDMI confidence with $v_{prior} = -2$, 0 or 2. The PDMI confidence intervals with $v_{prior} = 7$ are approximately as short as the ML confidence intervals, but have slightly lower coverage.

# 7 CONCLUSION

With our new confidence intervals, ML performs at least as well as MI in small samples. The new ML confidence intervals have close to nominal coverage and are shorter than MI confidence intervals with the same coverage. Our ML confidence intervals are easy to implement in SEM software, and we recommend that vendors make them part of the standard output. In addition, our results show that ML point estimates can be less biased and more precise than MI estimates in small samples. Previous results show that ML estimates are more precise than MI estimates in large samples as well (Wang & Robins, 1998).

The quality of estimates is not the only advantage of ML over MI. ML is also easier to use. MI runs much more slowly than ML, and MI requires the user to store and analyze multiple imputed copies of the data. Storage and runtime are not major concerns in small samples, but can be serious issues in large samples, particularly if the analysis runs slowly on each imputed dataset (von Hippel, 2005). In addition, MI often requires the user to specify separate models for the imputation and analysis, to ensure that assumptions made by the imputation model are not biasing the results of the analysis (Meng, 1994; Schafer, 1997). This can be a tricky issue, and can crop up in unexpected ways—e.g., when variables interact, or are transformed to reduce skew (von Hippel, 2009, 2013a). Finally, MI sometimes requires the user to make thorny decisions about Markov Chain Monte Carlo, an iterative method which is commonly used to generate posterior draws. None of these issues come up with ML.

Our general recommendation is to prefer ML (with the new confidence intervals) over MI whenever both options are available. But this does not mean than MI should never be used. There are common situations when ML is not an option. ML has rather strict requirements. ML requires a parametric model that can be estimated by maximum likelihood and used to derive the parameters of the analytic model. The analytic model need not be identical to the model that is used to maximize the likelihood, but it must be possible to derive the parameters of the analytic model from the parameters of the model that is used to maximize the likelihood (Savalei & Bentler, 2009; Yuan & Lu, 2008).

MI is more flexible. While the most popular flavor of MI has the same requirements as ML (Schafer, 1997), there are flavors of MI that do not require that the parameters of the analytic model can be derived from the parameters of the imputation model. MI can use very different parametric models for imputation and analysis (Raghunathan, Lepkowski, Van Hoewyk, & Solenberger, 2001; van Buuren & Oudshoorn, 1999). MI can also impute values nonparametrically, using a "hot-deck" procedure that replaces the missing values by sampling from the observed values (Andridge & Little, 2010). The flexibility of MI can have both advantages and disadvantages. The advantage is that the imputed values can look very realistic (Demirtas & Hedeker, 2008; He & Raghunathan, 2012). The potential disadvantage is that bias can be introduced by unnoticed conflicts between the assumptions of the imputation model and the assumptions of the analytic model (e.g., von Hippel, 2009, 2013a).

An advantage of MI is that it can be used when we plan to analyze the data using techniques that are not based on the likelihood, such as quantile regression or the generalized method of moments. Finally, MI is more widely implemented in software. Users who are not using SEM software—for example, a user running the LOGISTIC procedure in SAS—will often find that MI is available but ML is not.

When MI must be used, our results suggest several ways improve MI estimates. First, formula (27) offers a way to improve the degrees of freedom estimate that is used for PDMI confidence intervals. In addition, section 5.2 shows that PDMI point estimates can be improved by using a Bayesian prior with $2 \leq \nu_{prior} \leq 7$ degrees of freedom rather than the usual Jeffreys prior of $\nu_{prior} = 0$. Software should permit users to choose options in the range $0 \leq \nu_{prior} \leq 7$ perhaps with $\nu_{prior} = 2$ as a default. The option $\nu_{prior} = -2$, currently available in Stata (which calls it the uniform prior), produces such poor estimates that it should never be used.

For simplicity's sake, our presentation focused on bivariate normal data, but some of our conclusions apply more broadly. Our degrees of freedom formula (31) was derived from general principles and does not assume that the data are normally distributed. The small-sample biases that we observed under PDMI also have a rather general source: variation in the posterior draws adds to variation in the imputed variable. The biases introduced by this extra variation are not limited to the bivariate normal case. Indeed, similar biases have been observed in a simulation with 5 multivariate normal variables; that simulation also found that the biases were smaller under ML than under PDMI (Yuan et al., 2012, Table A10).

In non-normal data, the biases that we have observed might actually be worse. In normal data, the mean and variance estimates are independent, which is why our simulation found that a mean can be unbiased even if the variance is biased, or a regression slope can be unbiased if the residual variance is biased. Such examples would not be possible in binomial or Poisson data where the mean and variance are not independent parameters. In those settings, if the variance is biased, it seems inevitable that the mean and regression slopes will be biased as well. However, this prediction should be tested in future work.

# TABLES

Table 1. Expectations (standard errors) of estimators

|  |  |  | Estimands |  |  |  |  |  |  |  |  |
|---|---|---|---|---|---|---|---|---|---|---|---|
|  |  |  | Regression of $Y$ on $X$ | | | Moments of $Y$ | | | Regression of $X$ on $Y$ | | |
| Estimator | n | Pattern | $\alpha_{Y.X}$ | $\beta_{Y.X}$ | $\sigma^2_{Y.X}$ | $\mu_Y$ | $\sigma^2_Y$ | $\sigma_{XY}$ | $\alpha_{X.Y}$ | $\beta_{X.Y}$ | $\sigma^2_{X.Y}$ |
| ML | 25 | MXN | .00 | .50 | .62 | .00 | 1.09 | .48 | .10 | .41 | .63 |
|  |  |  | (.46) | (.51) | (.28) | (.50) | (.78) | (.48) | (.26) | (.39) | (.29) |
|  |  | MCAR | .00 | .50 | .62 | .00 | .94 | .48 | .00 | .52 | .65 |
|  |  |  | (.27) | (.29) | (.28) | (.28) | (.41) | (.30) | (.22) | (.29) | (.25) |
|  | 100 | MXN | .00 | .50 | .72 | .00 | 1.01 | .50 | .02 | .48 | .73 |
|  |  |  | (.21) | (.21) | (.15) | (.22) | (.27) | (.22) | (.12) | (.16) | (.17) |
|  |  | MCAR | .00 | .50 | .72 | .00 | .98 | .50 | .00 | .51 | .73 |
|  |  |  | (.12) | (.13) | (.15) | (.13) | (.20) | (.14) | (.10) | (.12) | (.13) |
| MLMI | 25 | MXN | .00 | .50 | .65 | .00 | 1.12 | .50 | .11 | .42 | .67 |
|  |  |  | (.46) | (.51) | (.30) | (.51) | (.81) | (.50) | (.26) | (.40) | (.31) |
|  |  | MCAR | .00 | .50 | .65 | .00 | .96 | .50 | .00 | .53 | .70 |
|  |  |  | (.27) | (.30) | (.30) | (.28) | (.44) | (.32) | (.22) | (.30) | (.27) |
|  | 100 | MXN | .00 | .50 | .73 | .00 | 1.02 | .50 | .02 | .48 | .74 |
|  |  |  | (.21) | (.21) | (.15) | (.22) | (.27) | (.22) | (.12) | (.16) | (.17) |
|  |  | MCAR | .00 | .50 | .73 | .00 | .99 | .50 | .00 | .51 | .74 |
|  |  |  | (.13) | (.13) | (.15) | (.14) | (.20) | (.15) | (.10) | (.12) | (.13) |
| PDMI | 25 | MXN | .00 | .50 | .87 | .00 | 1.58 | .50 | .17 | .32 | .67 |
|  |  |  | (.51) | (.56) | (.84) | (.56) | (1.82) | (.55) | (.22) | (.33) | (.28) |
|  |  | MCAR | .00 | .50 | .88 | .00 | 1.24 | .50 | .00 | .45 | .73 |
|  |  |  | (.29) | (.32) | (2.68) | (.30) | (2.98) | (.34) | (.21) | (.27) | (.26) |
|  | 100 | MXN | .00 | .50 | .77 | .00 | 1.09 | .50 | .04 | .44 | .74 |
|  |  |  | (.23) | (.23) | (.17) | (.23) | (.30) | (.24) | (.12) | (.16) | (.17) |
|  |  | MCAR | .00 | .50 | .77 | .00 | 1.03 | .50 | .00 | .49 | .75 |
|  |  |  | (.13) | (.13) | (.17) | (.14) | (.22) | (.15) | (.10) | (.12) | (.13) |
| Parameter values | | | 0 | .5 | .75 | 0 | 1 | .5 | 0 | .5 | .75 |

*Note*. In this table the PDMI estimates use $\nu_{prior} = 0$.

Table 2. Expectations (standard errors) of PDMI estimators with different values of $\nu_{prior}$

| | | | Estimands | | | | | | | | |
|---|---|---|---|---|---|---|---|---|---|---|---|
| | | | Regression of $Y$ on $X$ | | | Moments of $Y$ | | | Regression of $X$ on $Y$ | | |
| $\nu_{prior}$ | n | Pattern | $\alpha_{Y.X}$ | $\beta_{Y.X}$ | $\sigma^2_{Y.X}$ | $\mu_Y$ | $\sigma^2_Y$ | $\sigma_{XY}$ | $\alpha_{X.Y}$ | $\beta_{X.Y}$ | $\sigma^2_{X.Y}$ |
| –2 | 25 | MXN | .00 | .50 | 4.22 | .00 | 6.10 | .50 | .17 | .29 | .68 |
| | | | (.68) | (.82) | (678) | (.93) | (731) | (.72) | (.21) | (.30) | (.27) |
| | | MCAR | .00 | .50 | 205* | .00 | 198* | .50 | .00 | .41 | .75 |
| | | | (1.82) | (.64) | (76,986)* | (1.80) | (73,830)* | (.64) | (.21) | (.25) | (.26) |
| | 100 | MXN | .00 | .50 | .79 | .00 | 1.11 | .50 | .04 | .43 | .74 |
| | | | (.23) | (.23) | (.17) | (.24) | (.31) | (.24) | (.12) | (.16) | (.17) |
| | | MCAR | .00 | .50 | .79 | .00 | 1.05 | .50 | .00 | .48 | .75 |
| | | | (.13) | (.13) | (.17) | (.14) | (.22) | (.15) | (.10) | (.12) | (.13) |
| 0 | 25 | MXN | .00 | .50 | .87 | .00 | 1.58 | .50 | .17 | .32 | .67 |
| | | | (.51) | (.56) | (.84) | (.56) | (1.82) | (.55) | (.22) | (.33) | (.28) |
| | | MCAR | .00 | .50 | .88 | .00 | 1.24 | .50 | .00 | .45 | .73 |
| | | | (.29) | (.32) | (2.68) | (.30) | (2.98) | (.34) | (.21) | (.27) | (.26) |
| | 100 | MXN | .00 | .50 | .77 | .00 | 1.09 | .50 | .04 | .44 | .74 |
| | | | (.23) | (.23) | (.17) | (.23) | (.30) | (.24) | (.12) | (.16) | (.17) |
| | | MCAR | .00 | .50 | .77 | .00 | 1.03 | .50 | .00 | .49 | .75 |
| | | | (.13) | (.13) | (.17) | (.14) | (.22) | (.15) | (.10) | (.12) | (.13) |
| 2 | 25 | MXN | .00 | .50 | .75 | .00 | 1.40 | .50 | .16 | .35 | .66 |
| | | | (.49) | (.55) | (.37) | (.54) | (1.03) | (.54) | (.23) | (.35) | (.28) |
| | | MCAR | .00 | .50 | .75 | .00 | 1.09 | .50 | .00 | .48 | .71 |
| | | | (.28) | (.31) | (.37) | (.29) | (.54) | (.33) | (.22) | (.29) | (.26) |
| | 100 | MXN | .00 | .50 | .75 | .00 | 1.07 | .50 | .04 | .45 | .74 |
| | | | (.23) | (.23) | (.16) | (.24) | (.30) | (.24) | (.12) | (.17) | (.17) |
| | | MCAR | .00 | .50 | .75 | .00 | 1.02 | .50 | .00 | .50 | .74 |
| | | | (.13) | (.13) | (.16) | (.14) | (.21) | (.15) | (.10) | (.12) | (.13) |
| 7 | 25 | MXN | .00 | .50 | .61 | .00 | 1.20 | .50 | .16 | .40 | .64 |
| | | | (.48) | (.53) | (.29) | (.53) | (.92) | (.53) | (.24) | (.39) | (.29) |
| | | MCAR | .00 | .50 | .61 | .00 | .94 | .50 | .00 | .54 | .68 |
| | | | (.28) | (.30) | (.28) | (.29) | (.47) | (.32) | (.22) | (.31) | (.26) |
| | 100 | MXN | .00 | .50 | .71 | .00 | 1.03 | .50 | .04 | .47 | .73 |
| | | | (.22) | (.23) | (.15) | (.23) | (.29) | (.24) | (.12) | (.17) | (.17) |
| | | MCAR | .00 | .50 | .71 | .00 | .98 | .50 | .00 | .51 | .74 |
| | | | (.13) | (.13) | (.15) | (.14) | (.21) | (.15) | (.10) | (.12) | (.13) |
| Parameter values | | | 0 | .5 | .75 | 0 | 1 | .5 | 0 | .5 | .75 |

* With $\nu_{prior} = -2$, the expectation or standard error of certain estimates can be undefined. In the table, these undefined values show up as very large numbers.

Table 3. Value (rank) of RMSE for different estimators

| Estimator | $v_{prior}$ | n | Pattern | Regression of Y on X | | | Moments of Y | | | Regression of X on Y | | |
|---|---|---|---|---|---|---|---|---|---|---|---|---|
| | | | | $\alpha_{Y,X}$ | $\beta_{Y,X}$ | $\sigma^2_{Y,X}$ | $\mu_Y$ | $\sigma^2_Y$ | $\sigma_{XY}$ | $\alpha_{X,Y}$ | $\beta_{X,Y}$ | $\sigma^2_{X,Y}$ |
| ML | | 25 | MXN | .457 (1) | .712 (1) | .684 (2) | .503 (1) | 1.335 (1) | .679 (1) | .280 (4) | .565 (5) | .695 (1) |
| | | | MCAR | .266 (1) | .580 (1) | .682 (2) | .278 (1) | 1.025 (1) | .567 (1) | .216 (4) | .594 (4) | .699 (1) |
| | | 100 | MXN | .209 (1) | .543 (1) | .736 (2) | .218 (1) | 1.046 (1) | .542 (1) | .124 (4) | .503 (5) | .746 (1) |
| | | | MCAR | .124 (1) | .516 (1) | .735 (2) | .133 (1) | 1.003 (2) | .516 (1) | .099 (1) | .519 (4) | .740 (1) |
| MLMI | | 25 | MXN | .460 (2) | .714 (2) | .716 (3) | .505 (2) | 1.381 (2) | .709 (2) | .285 (5) | .578 (6) | .742 (6) |
| | | | MCAR | .270 (2) | .582 (2) | .715 (3) | .282 (2) | 1.058 (3) | .593 (2) | .220 (5) | .607 (5) | .747 (3) |
| | | 100 | MXN | .211 (2) | .544 (2) | .744 (3) | .220 (2) | 1.054 (2) | .548 (2) | .125 (5) | .506 (6) | .758 (4) |
| | | | MCAR | .127 (2) | .517 (2) | .743 (3) | .136 (2) | 1.010 (3) | .522 (2) | .100 (4) | .522 (5) | .752 (3) |
| PDMI | -2 | 25 | MXN | .682 (6) | .957 (6) | 678 (6) | .926 (6) | 731 (6) | .873 (6) | .266 (1) | .420 (1) | .733 (5) |
| | | | MCAR | 1.819 (6) | .810 (6) | 76986* (6) | 1.801 (6) | 73830* (6) | .809 (6) | .207 (1) | .479 (1) | .790 (6) |
| | | 100 | MXN | .228 (6) | .551 (6) | .805 (6) | .237 (6) | 1.155 (6) | .555 (6) | .123 (2) | .462 (1) | .762 (6) |
| | | 25 | MCAR | .131 (6) | .518 (4) | .805 (6) | .139 (6) | 1.076 (6) | .523 (4) | .100 (2) | .494 (1) | .764 (6) |
| | | | MCAR | .682 (6) | .957 (6) | 678 (6) | .926 (6) | 731.552 (6) | .873 (6) | .266 (1) | .420 (1) | .733 (5) |
| PDMI | 0 | 25 | MXN | .506 (5) | .752 (5) | 1.208 (5) | .559 (5) | 2.414 (5) | .744 (5) | .272 (2) | .461 (2) | .721 (4) |
| | | | MCAR | .286 (5) | .592 (5) | 2.823 (5) | .296 (5) | 3.230 (5) | .602 (5) | .212 (2) | .526 (2) | .770 (5) |
| | | 100 | MXN | .226 (4) | .550 (4) | .786 (5) | .235 (4) | 1.133 (5) | .554 (4) | .123 (1) | .470 (2) | .759 (5) |
| | | 25 | MCAR | .130 (5) | .518 (6) | .785 (5) | .139 (5) | 1.057 (5) | .523 (5) | .100 (3) | .503 (2) | .760 (5) |
| | | | MCAR | .286 (5) | .592 (5) | 2.823 (5) | .296 (5) | 3.230 (5) | .602 (5) | .212 (2) | .526 (2) | .770 (5) |
| PDMI | 2 | 25 | MXN | .494 (4) | .742 (4) | .835 (4) | .544 (4) | 1.736 (4) | .735 (4) | .277 (3) | .495 (3) | .714 (3) |
| | | | MCAR | .280 (4) | .589 (4) | .832 (4) | .291 (4) | 1.220 (4) | .599 (4) | .216 (3) | .562 (3) | .755 (4) |
| | | 100 | MXN | .226 (5) | .551 (5) | .768 (4) | .235 (5) | 1.116 (4) | .555 (5) | .124 (3) | .478 (3) | .756 (3) |
| | | 25 | MCAR | .130 (4) | .518 (5) | .767 (4) | .138 (4) | 1.039 (4) | .523 (6) | .100 (5) | .511 (3) | .756 (4) |
| | | | MCAR | .494 (4) | .742 (4) | .835 (4) | .544 (4) | 1.736 (4) | .735 (4) | .277 (3) | .495 (3) | .714 (3) |
| PDMI | 7 | 25 | MXN | .481 (3) | .732 (3) | .673 (1) | .528 (3) | 1.507 (3) | .725 (3) | .289 (6) | .556 (4) | .700 (2) |
| | | | MCAR | .275 (3) | .586 (3) | .672 (1) | .286 (3) | 1.051 (2) | .596 (3) | .225 (6) | .627 (6) | .729 (2) |
| | | 100 | MXN | .224 (3) | .549 (3) | .731 (1) | .232 (3) | 1.074 (3) | .553 (3) | .125 (6) | .495 (4) | .749 (2) |
| | | 25 | MCAR | .130 (3) | .517 (3) | .730 (1) | .138 (3) | 1.001 (1) | .522 (3) | .101 (6) | .528 (6) | .748 (2) |
| | | | MCAR | .481 (3) | .732 (3) | .673 (1) | .528 (3) | 1.507 (3) | .725 (3) | .289 (6) | .556 (4) | .700 (2) |

*Note.* For each sample size and pattern, the estimator with the smallest RMSE is ranked (1), and the estimator with the largest RMSE is ranked (6).

* With $v_{prior} = -2$, the expectation or standard error of certain estimates can be undefined. In the table, these undefined values show up as very large number.

Table 4. Mean length (and coverage %) of nominal 95% confidence intervals, using the usual formulas

| Estimator | $\nu_{prior}$ | n | Pattern | Regression of Y on X | | | Moments of Y | | | Regression of X on Y | | |
|---|---|---|---|---|---|---|---|---|---|---|---|---|
| | | | | $\alpha_{Y.X}$ | $\beta_{Y.X}$ | $\sigma^2_{Y.X}$ | $\mu_Y$ | $\sigma^2_Y$ | $\sigma_{XY}$ | $\alpha_{X.Y}$ | $\beta_{X.Y}$ | $\sigma^2_{X.Y}$ |
| ML | | 25 | MXN | 1.5 | 1.6 | 1.0 | 1.7 | 2.4 | 1.7 | 1.1 | 1.3 | 1.2 |
| | | | | (89%) | (90%) | (83%) | (91%) | (90%) | (92%) | (90%) | (90%) | (87%) |
| | | | MCAR | 0.9 | 1.0 | 1.0 | 1.0 | 1.5 | 1.1 | 0.8 | 1.0 | 0.9 |
| | | | | (90%) | (89%) | (82%) | (91%) | (87%) | (90%) | (92%) | (90%) | (87%) |
| | | 100 | MXN | 0.8 | 0.8 | 0.6 | 0.8 | 1.0 | 0.8 | 0.5 | 0.6 | 0.6 |
| | | | | (94%) | (94%) | (92%) | (94%) | (93%) | (94%) | (94%) | (94%) | (92%) |
| | | | MCAR | 0.5 | 0.5 | 0.6 | 0.5 | 0.8 | 0.6 | 0.4 | 0.4 | 0.5 |
| | | | | (94%) | (94%) | (92%) | (94%) | (93%) | (94%) | (94%) | (94%) | (93%) |
| PDMI | -2 | 25 | MXN | 1.9E+10 | 1.3E+11 | 3.4E+13 | 60.3 | 1.6E+181 | 4.8E+09 | 1.6 | 3.0E+04 | 3.3 |
| | | | | (97%) | (98%) | (97%) | (97%) | (98%) | (98%) | (95%) | (96%) | (96%) |
| | | | MCAR | 2.4 | 65.3 | 4.8E+16 | 2.2 | 2.9E+16 | 40.3 | 0.9 | 1.6 | 3.1 |
| | | | | (97%) | (97%) | (97%) | (97%) | (97%) | (97%) | (96%) | (96%) | (96%) |
| | | 100 | MXN | 1.3 | 1.3 | 0.8 | 1.2 | 1.5 | 1.2 | 0.6 | 0.8 | 0.8 |
| | | | | (95%) | (95%) | (95%) | (95%) | (95%) | (95%) | (96%) | (96%) | (94%) |
| | | | MCAR | 0.6 | 0.6 | 0.8 | 0.6 | 1.0 | 0.7 | 0.4 | 0.5 | 0.5 |
| | | | | (95%) | (95%) | (95%) | (95%) | (95%) | (96%) | (96%) | (95%) | (95%) |
| PDMI | 0 | 25 | MXN | 2.5E+10 | 2.7E+06 | 807.5 | 9.4 | 1.1E+05 | 4.1 | 1.6 | 154.8 | 2.7 |
| | | | | (97%) | (97%) | (94%) | (96%) | (97%) | (97%) | (94%) | (96%) | (94%) |
| | | | MCAR | 1.9 | 1.0E+13 | 3.1E+08 | 1.7 | 1.8E+08 | 1.8 | 0.9 | 1.4 | 1.4 |
| | | | | (96%) | (95%) | (94%) | (96%) | (96%) | (96%) | (96%) | (95%) | (94%) |
| | | 100 | MXN | 1.2 | 1.3 | 0.7 | 1.2 | 1.5 | 1.2 | 0.6 | 0.8 | 0.8 |
| | | | | (95%) | (95%) | (94%) | (95%) | (94%) | (95%) | (96%) | (96%) | (94%) |
| | | | MCAR | 0.6 | 0.6 | 0.7 | 0.6 | 1.0 | 0.6 | 0.4 | 0.5 | 0.5 |
| | | | | (95%) | (95%) | (94%) | (96%) | (95%) | (95%) | (95%) | (95%) | (95%) |
| PDMI | 2 | 25 | MXN | 4.4E+07 | 1.8E+07 | 9.0E+04 | 53.7 | 3.3E+08 | 3.6 | 1.6 | 106.6 | 3.7 |
| | | | | (96%) | (96%) | (90%) | (95%) | (96%) | (96%) | (93%) | (95%) | (93%) |
| | | | MCAR | 1.5 | 60.6 | 1.4E+10 | 1.4 | 9.8E+09 | 1.6 | 0.9 | 1.4 | 1.2 |
| | | | | (95%) | (94%) | (91%) | (95%) | (94%) | (95%) | (95%) | (94%) | (92%) |
| | | 100 | MXN | 1.2 | 1.2 | 0.7 | 1.2 | 1.4 | 1.2 | 0.6 | 0.8 | 0.8 |
| | | | | (95%) | (94%) | (93%) | (95%) | (94%) | (95%) | (95%) | (95%) | (94%) |
| | | | MCAR | 0.6 | 0.6 | 0.7 | 0.6 | 0.9 | 0.6 | 0.4 | 0.5 | 0.5 |
| | | | | (95%) | (94%) | (94%) | (95%) | (95%) | (95%) | (95%) | (94%) | (94%) |
| PDMI | 7 | 25 | MXN | 1.0E+16 | 7.3E+15 | 1.3 | 4.8 | 13.2 | 2.8 | 1.5 | 63.1 | 2.0 |
| | | | | (94%) | (94%) | (79%) | (93%) | (92%) | (95%) | (91%) | (92%) | (90%) |
| | | | MCAR | 1.2 | 58.8 | 1.1 | 1.2 | 2.0 | 1.4 | 0.9 | 1.2 | 1.1 |
| | | | | (92%) | (91%) | (79%) | (93%) | (88%) | (93%) | (94%) | (90%) | (88%) |
| | | 100 | MXN | 1.2 | 1.2 | 0.6 | 1.1 | 1.3 | 1.1 | 0.6 | 0.8 | 0.8 |
| | | | | (94%) | (94%) | (91%) | (94%) | (93%) | (94%) | (95%) | (94%) | (93%) |
| | | | MCAR | 0.6 | 0.6 | 0.6 | 0.6 | 0.9 | 0.6 | 0.4 | 0.5 | 0.5 |
| | | | | (94%) | (93%) | (90%) | (95%) | (92%) | (95%) | (95%) | (93%) | (93%) |

*Note.* In some cells, very large estimates indicate that the mean length of the confidence interval is undefined.

Table 5. Mean length (and coverage %) of nominal 95% confidence intervals, using our improved formulas

| Estimator | $\nu_{prior}$ | n | Pattern | Regression of Y on X | | | Moments of Y | | | Regression of X on Y | | |
|---|---|---|---|---|---|---|---|---|---|---|---|---|
| | | | | $\alpha_{Y.X}$ | $\beta_{Y.X}$ | $\sigma^2_{Y.X}$ | $\mu_Y$ | $\sigma^2_Y$ | $\sigma_{XY}$ | $\alpha_{X.Y}$ | $\beta_{X.Y}$ | $\sigma^2_{X.Y}$ |
| ML | | 25 | MXN | 2.4 | 2.6 | 1.2 | 2.4 | 3.4 | 2.3 | 1.4 | 1.8 | 1.5 |
| | | | | (98%) | (98%) | (87%) | (97%) | (94%) | (98%) | (93%) | (94%) | (93%) |
| | | | MCAR | 1.1 | 1.2 | 1.2 | 1.1 | 1.8 | 1.2 | 0.8 | 1.1 | 1.0 |
| | | | | (93%) | (93%) | (86%) | (94%) | (91%) | (93%) | (94%) | (93%) | (90%) |
| | | 100 | MXN | 0.9 | 0.9 | 0.6 | 0.9 | 1.1 | 0.9 | 0.5 | 0.6 | 0.7 |
| | | | | (96%) | (96%) | (93%) | (96%) | (93%) | (96%) | (95%) | (94%) | (93%) |
| | | | MCAR | 0.5 | 0.5 | 0.6 | 0.5 | 0.8 | 0.6 | 0.4 | 0.5 | 0.5 |
| | | | | (95%) | (94%) | (93%) | (95%) | (94%) | (95%) | (95%) | (94%) | (94%) |
| PDMI | -2 | 25 | MXN | 3.7 | 3.9 | 7.0 | 4.0 | 26.5 | 3.9 | 1.5 | 1.9 | 1.7 |
| | | | | (97%) | (97%) | (96%) | (97%) | (98%) | (98%) | (95%) | (96%) | (96%) |
| | | | MCAR | 2.0 | 2.2 | 166.5 | 2.0 | 403.6 | 2.3 | 0.9 | 1.3 | 1.2 |
| | | | | (97%) | (97%) | (97%) | (97%) | (97%) | (97%) | (96%) | (96%) | (96%) |
| | | 100 | MXN | 1.2 | 1.3 | 0.8 | 1.2 | 1.5 | 1.2 | 0.6 | 0.8 | 0.8 |
| | | | | (95%) | (95%) | (95%) | (95%) | (95%) | (95%) | (96%) | (96%) | (94%) |
| | | | MCAR | 0.6 | 0.6 | 0.8 | 0.6 | 1.0 | 0.7 | 0.4 | 0.5 | 0.5 |
| | | | | (95%) | (95%) | (95%) | (95%) | (95%) | (96%) | (96%) | (95%) | (95%) |
| PDMI | 0 | 25 | MXN | 3.1 | 3.3 | 2.5 | 3.2 | 6.4 | 3.2 | 1.5 | 1.9 | 1.6 |
| | | | | (96%) | (96%) | (94%) | (96%) | (97%) | (97%) | (94%) | (95%) | (94%) |
| | | | MCAR | 1.5 | 1.7 | 2.6 | 1.6 | 3.6 | 1.7 | 0.9 | 1.3 | 1.2 |
| | | | | (96%) | (95%) | (94%) | (96%) | (96%) | (96%) | (96%) | (95%) | (94%) |
| | | 100 | MXN | 1.2 | 1.2 | 0.7 | 1.2 | 1.5 | 1.2 | 0.6 | 0.8 | 0.8 |
| | | | | (95%) | (95%) | (94%) | (95%) | (94%) | (95%) | (96%) | (96%) | (94%) |
| | | | MCAR | 0.6 | 0.6 | 0.7 | 0.6 | 1.0 | 0.6 | 0.4 | 0.5 | 0.5 |
| | | | | (95%) | (95%) | (94%) | (96%) | (95%) | (95%) | (95%) | (95%) | (95%) |
| PDMI | 2 | 25 | MXN | 2.8 | 2.9 | 1.8 | 2.9 | 5.2 | 2.9 | 1.4 | 1.9 | 1.5 |
| | | | | (95%) | (95%) | (90%) | (95%) | (95%) | (96%) | (93%) | (95%) | (93%) |
| | | | MCAR | 1.4 | 1.5 | 1.8 | 1.4 | 2.7 | 1.5 | 0.9 | 1.3 | 1.1 |
| | | | | (95%) | (94%) | (90%) | (95%) | (94%) | (95%) | (95%) | (94%) | (92%) |
| | | 100 | MXN | 1.2 | 1.2 | 0.7 | 1.2 | 1.4 | 1.2 | 0.6 | 0.8 | 0.8 |
| | | | | (95%) | (94%) | (93%) | (95%) | (94%) | (95%) | (95%) | (95%) | (94%) |
| | | | MCAR | 0.6 | 0.6 | 0.7 | 0.6 | 0.9 | 0.6 | 0.4 | 0.5 | 0.5 |
| | | | | (95%) | (94%) | (94%) | (95%) | (95%) | (95%) | (95%) | (94%) | (94%) |
| PDMI | 7 | 25 | MXN | 2.3 | 2.4 | 1.1 | 2.4 | 3.6 | 2.4 | 1.4 | 1.8 | 1.4 |
| | | | | (93%) | (93%) | (79%) | (93%) | (92%) | (95%) | (91%) | (92%) | (89%) |
| | | | MCAR | 1.1 | 1.2 | 1.1 | 1.2 | 1.9 | 1.3 | 0.9 | 1.2 | 1.0 |
| | | | | (91%) | (91%) | (79%) | (93%) | (88%) | (93%) | (94%) | (90%) | (88%) |
| | | 100 | MXN | 1.1 | 1.2 | 0.6 | 1.1 | 1.3 | 1.1 | 0.6 | 0.8 | 0.8 |
| | | | | (94%) | (94%) | (91%) | (94%) | (93%) | (94%) | (95%) | (94%) | (93%) |
| | | | MCAR | 0.6 | 0.6 | 0.6 | 0.6 | 0.9 | 0.6 | 0.4 | 0.5 | 0.5 |
| | | | | (94%) | (93%) | (90%) | (95%) | (92%) | (95%) | (95%) | (93%) | (93%) |

# APPENDIX: BIAS CALCULATIONS

In this Appendix we approximate the bias of $\hat{\sigma}^2_{Y,MLSI}$ and $\hat{\sigma}^2_{Y,PDSI}$.

Under PDSI or MLSI, the distribution of the partly observed, parly imputed variable $Y_{obsSI}$ can be written as

$$Y_{obsSI} = \alpha + M\Delta\hat{\alpha}_{Y.X} + \beta X + XM\Delta\hat{\beta}_{Y.X} + e_{obsSI}, \qquad (35).$$
$$\text{where } e_{obsSI} \sim N(0, \sigma^2_{Y.X} + M\Delta\hat{\sigma}^2_{Y.X})$$

where $\Delta\hat{\alpha}_{Y.X}$, $\Delta\hat{\beta}_{Y.X}$, and $\Delta\hat{\sigma}^2_{Y.X}$ are the errors of the ML or PD estimates that are used in the imputation model. There are a lot of random variables in this equation, but the situation becomes more tractable if we accept the approximation that $X$, $M$, $MX$ are fixed with covariance matrix $\Sigma_{M,X,MX}$.

Then the SI estimator

$$\hat{\sigma}^2_{Y,SI} = \frac{1}{n-1}\sum_{i=1}^{n}\left(Y_{obsSI,i} - \hat{\mu}_{Y,SI}\right)^2 \qquad (36)$$

has conditional expectation

$$E(\hat{\sigma}^2_{Y,SI}|\hat{\alpha}_{Y.X}, \hat{\beta}_{Y.X}, \hat{\sigma}^2_{Y.X}) = V(Y_{obsSI}|\hat{\alpha}_{Y.X}, \hat{\beta}_{Y.X}, \hat{\sigma}^2_{Y.X})$$
$$= \begin{bmatrix}\Delta\hat{\alpha}_{Y.X}\\ \beta \\ \Delta\hat{\beta}_{Y.X}\end{bmatrix}^T \Sigma_{M,X,MX} \begin{bmatrix}\Delta\hat{\alpha}_{Y.X}\\ \beta \\ \Delta\hat{\beta}_{Y.X}\end{bmatrix} + \sigma^2_{Y.X} + p\Delta\hat{\sigma}^2_{Y.X} \qquad (37).$$

Breaking this into components, we see that the first component is a quadratic form whose expectation can be calculated using a standard matrix identity (Seber & Lee, 2003, pp. 9–12),

$$E\left(\begin{bmatrix}\Delta\hat{\alpha}_{Y.X}\\ \beta \\ \Delta\hat{\beta}_{Y.X}\end{bmatrix}^T \Sigma_{M,X,MX} \begin{bmatrix}\Delta\hat{\alpha}_{Y.X}\\ \beta \\ \Delta\hat{\beta}_{Y.X}\end{bmatrix}\right) = tr\left(\Sigma_{M,X,MX} V\begin{bmatrix}\Delta\hat{\alpha}_{Y.X}\\ \beta \\ \Delta\hat{\beta}_{Y.X}\end{bmatrix}\right) + \begin{bmatrix}0\\ \beta \\ 0\end{bmatrix}^T \Sigma_{M,X,MX} \begin{bmatrix}0\\ \beta \\ 0\end{bmatrix}^T \qquad (38),$$

$$= tr(\Sigma_{MXM}\Sigma_{\hat{\alpha}\hat{\beta}_{Y.X}}) + \beta^2\sigma_X^2$$

where $\Sigma_{MXM}$ is the covariance matrix of (M,XM).

The remaining components have expectation

$$\sigma_{Y.X}^2 + p\Delta\hat{\sigma}_{Y.X}^2 = \sigma_{Y.X}^2 + pE(\Delta\hat{\sigma}_{Y.X}^2)$$
$$= \sigma_{Y.X}^2 + pBias(\hat{\sigma}_{Y.X}^2)$$
(39),

where $Bias(\hat{\sigma}_{Y.X}^2)$ is the bias of $\hat{\sigma}_{Y.X}^2$ as an estimator of $\sigma_{Y.X}^2$.

Therefore the unconditional expectation of $\hat{\sigma}_{Y,SI}^2$ is

$$E(\hat{\sigma}_{Y,SI}^2) = tr(\Sigma_{MXM}\Sigma_{\hat{\alpha}\hat{\beta}_{Y.X}}) + \beta^2\sigma_X^2 + \sigma_{Y.X}^2 + pBias(\hat{\sigma}_{Y.X}^2)$$

And the bias of $\hat{\sigma}_{Y,SI}^2$ as an estimator of $\sigma_{Y.X}^2$ is

$$Bias(\hat{\sigma}_{Y,SI}^2) = E(\hat{\sigma}_{Y,SI}^2) - \sigma_Y^2 = E(\hat{\sigma}_{Y,SI}^2) - \beta^2\sigma_X^2 - \sigma_{Y.X}^2$$
$$= tr(\Sigma_{MXM}\Sigma_{\hat{\alpha}\hat{\beta}_{Y.X}}) + pBias(\hat{\sigma}_{Y.X,SI}^2)$$

This justifies the expressions in (15) and (20).

# ONLINE APPENDIX:
# SMALL SAMPLE ML CONFIDENCE INTERVALS IN SAS

In this online Appendix we illustrate how to calculate the fraction of missing information, degrees of freedom, and nominal 95% *t*-based confidence intervals according to section 6.2. We use the CALIS procedure in SAS version 9.3. In developing this code, we uncovered an error in the CALIS procedure, which SAS corrected with Hot Fix I41012. Before the Hot Fix, the CALIS procedure calculated standard errors using the expected information matrix, which only gives valid results if values are MCAR (Yuan & Bentler, 2007). Hot Fix I41012 switched to the observed information matrix, which also gives valid results if values are MAR. Be sure the Hot Fix is installed before running the SAS code below.

```sas
%let n = 25;

/* Make incomplete data */
data incomplete;
 do i = 1 to &n;
  x = rannor(1);
  y = .5 * x + sqrt(1-.5**2) * rannor(1);
  if x < 0 then y = .;
  output;
 end;
run;

/* Use full information ML to estimate the mean and covariance matrix */
ods trace on / listing;
proc calis data=incomplete method=fiml;
 mean x y;
 var x y ;
 ods output LINEQSMeans=mean LINEQSVarExog=var LINEQSCovExog=cov;
run;

/* Reformat the mean and covariance matrix in format expected for CALIS
input. (This is cumbersome.) */
data MeanVarCov ( keep=Var1-Var2 _type_ Estimate rename=(Var1=_name_));
 set mean (in=_mean_) var (in=_var_) cov (in=_cov_);
 if _mean_ then _type_="MEAN";
 if (_var_ or _cov_) then _type_="COV";
 if not _cov_ then Var1=Variable;
 if _var_ then Var2=Var1;
run;
proc sort data=MeanVarCov;
 by _type_ Var2;
run;
proc transpose data=MeanVarCov out=MeanVarCov_t (drop=_name_
rename=(Var2=_name_));
 var estimate;
 by _type_ Var2;
run;
data N;
 _type_="N"; x=25; y=25;
 output;
run;
data MeanVarCovN (type=cov);
 set MeanVarCov_t N;
run;

/* Obtain full information ML estimate and SE of the regression slope from
the incomplete data */
proc calis data=incomplete method=fiml ;
 path y <- x;
 ods output PathList=PathList_incomplete (keep=parameter estimate stderr);
run;
proc print data=PathList_incomplete;
run;

/* Estimate SE that would have been obtained from complete data */
proc calis data=MeanVarCovN method=ml;
 path y <- x;
 ods output PathList=PathList_complete (keep=parameter estimate stderr);
run;
proc print data=PathList_complete;
run;

/* Calculate fraction of incomplete information, df, and nominal 95%
confidence interval */
data PathList;
 merge PathList_incomplete (rename=(stderr=stderr_incomplete))
```

```
            PathList_complete (rename=(stderr=stderr_complete)) ;
 by parameter;
 fraction_incomplete = 1 - (stderr_complete / stderr_incomplete)**2;
 df_comp = &n - 2;
 df = max (3, df_comp * (1-fraction_incomplete) * (df_comp+1) /
(df_comp+3));
 lcl = estimate - tinv(.975,df) * stderr_incomplete;
 ucl = estimate + tinv(.975,df) * stderr_incomplete;
run;
proc print data=PathList;
run;
```